\documentclass[onecolumn,11pt,aps,nofootinbib,superscriptaddress,tightenlines]{revtex4}
\usepackage{graphics,graphicx,psfrag,color,float, hyperref}
\usepackage{epsfig}
\usepackage{amsmath}
\usepackage{amssymb}
\usepackage{float}

\usepackage{amsmath}
\usepackage{amssymb}
\usepackage{amsfonts}
\usepackage{amsthm, hyperref}
\usepackage {times}
\usepackage{braket}
\usepackage{amsfonts}

\newcommand{\beq}{\begin{equation}}
\newcommand{\enq}{\end{equation}}
\newcommand{\bel}{\begin{lemma}}
\newcommand{\enl}{\end{lemma}}
\newcommand{\bet}{\begin{theorem}}
\newcommand{\ent}{\end{theorem}}

\newcommand{\lda}{\lambda}

\newcommand*{\cA}{\mathcal{A}}

\newcommand*{\bP}{\mathbf{P}}

\newcommand{\gma}{\gamma}

\newtheorem{definition}{Definition}

\newtheorem{theorem}{Theorem}
\newtheorem{lemma}{Lemma}

\begin {document}

\title{Power Control in Multiuser Mulicarrier Wireless Data Networks}

\author{Naqueeb Ahmad Warsi}
\email{naqueeb@tifr.res.in}
\affiliation{School of Technology and Computer Science,
Tata Institute of Fundamental Research (TIFR), Mumbai 400 005, India}


\begin{abstract}
A game-theoretic model is presented to study the management of transmission power in a wireless data network. We propose a power game 
for a multiuser multicarrier setting where all the users are assumed to transmit at equal rate. At equilibrium, each user is shown to transmit over a 
single carrier, as in [Mehskati et al., 2006]. We derive the necessary conditions on the path gains when the Nash equilibrium point exists.
We further prove the existence of the Nash equilibrium point using the concept of \emph {locally gross direction preserving map}. A greedy algorithm is 
proposed and its correctness is established, where each user acts selfishly to achieve the Nash equilibrium point.
\end{abstract}

\maketitle

\section{Introduction}
 {\huge{E}}ffective radio resource management is essential to promote the quality and efficiency of a wireless system. One of the major components of radio resource management is
power control, the subject of study of this paper. The principal purpose of power control is to provide each signal with adequate quality
without causing unnecessary interference to other signals. Another goal is to minimize the battery drain in the terminals. To formulate the power
control problem for multiuser multicarrier setting, we use the terms of economics where QoS (quality of service) objective is referred to as utility function.

Game theory has been widely used in the recent past to study the resource allocation problem in multiple access wireless systems (see Refs.\cite{Fahad,Meshkati,Mandayam,Samson}).
In Ref. \cite{Meshkati}, the authors study the maximization of utility under the average source rate and transmission delay constraints. 
In Ref. \cite{Mandayam}, the authors introduce pricing on transmit power to obtain Pareto improvement of the noncooperative power control game. 
In Ref. \cite{Samson}, a decentralized power allocation algorithm is proposed using concepts of game theory and random matrix theory 
for the case of fading MIMO multiple access channel.

To the best of our knowledge, the only paper where a power control game for a multiuser multicarrier
data network was analyzed is Ref. \cite{Fahad}, where the authors applied the framework developed in Ref. \cite{Goodman} to the multiuser multicarrier DS-CDMA 
data networks. However in Ref. \cite{Fahad}, the authors assume that there is no co-channel interference.
The authors then derive the optimal transmission strategy for each user. Furthermore the necessary conditions on the channel gains are derived when Nash equilibrium exists.

Compared with the previous work we let go the assumption of \emph{zero co-channel interference}. Furthermore, under certain assumptions (see \eqref{assumption} in Section\ref{direction preserving}
), we prove the existence of a Nash 
equilibrium point for the proposed power control game by using the concept of locally gross direction preserving map Ref. \cite{Hearing}.

This paper is organized as follows. In Section \ref{basic}, we  discuss some basic definitions and results from the fixed-point theory. In Section \ref{system model}, we give the system model for the multiuser
multicarrier multiple access data network, where we assume that there are $N$ transmitter-receiver pairs and develop a utility function that represents the QoS of data users. 
In Section \ref{formulation}, we give a game-theoretical formulation for the power control in multiuser multicarrier data network.
In Section \ref{best}, we discuss about the Nash equilibrium for the proposed game. In Section \ref{direction preserving}, we prove the existence of a Nash equilibrium point for the proposed game
by using the concept of locally gross direction preserving map. In Section \ref{algo}, we propose a greedy algorithm where all the users choose their transmit
power selfishly to achieve the Nash equilibrium point. In Section \ref{simulation} we discuss simulation results.
Conslcusions are given in Section \ref{con}.

\section {Some definitions and standard results from fixed-point theory}
\label{basic}
In this section  we recall and unify some standard definitions and results from the fixed-point theory \cite{Agarwal}.
\subsection{Utility function}
A utility function maps the element of the action set $\cal{A}$ to real numbers, i.e., $U:\cal{A}\rightarrow\mathbb{R}$, if  
$\forall ~ i,j \in A$, $i$ is at least as preferred as $j$ if and only if $U(i)\geq U(j)$. Informally, a utility function can be described
as the amount of satisfaction an agent receives as a result of the action. In wireless data networks the term utility is closely 
related to QoS objective. One of the most important QoS objectives in wireless date network is the low probability of error. The probability
of error is a function of SINR (signal to interference and noise ratio), $\gma$, hence $\gma$ is important in a wireless network. The probability of error approaches 0, for a
high $\gma$ and is very high for a small $\gma$. An important factor in the utility of all data systems is power consumption.
The level of satisfaction for someone using battery powered devices depends on how often he has to replace his battery; the battery life
is inversely proportional to the power drain on the batteries. Thus, the utility function depends on both $\gma$ and the transmitted power.

\subsection{Existence of a fixed Point}

Let $\lda:\cal{A} \mapsto \cal{A}$ be any mapping from a subset $\cal{A}$ $\subseteq \mathbb{R}^{m}$ to itself. One can associate $\lambda$ to 
a dynamical system described by the following discrete time equation:
\beq
\label{dynamical}
 \bP(n+1)=\lda(\bP(n)), n\in \mathbb{N}_{+},
\enq

\noindent where $\bP(n)\in \cal{A}$ is the vector of the state variables of the system at discrete time n. The equilibria of the system,
if they exist, are the vectors $\mathbf{P}^{*}$ resulting as a solution of  $\mathbf{P}^{*} = \lda(\mathbf{P}^{*})$, i.e., the fixed-points of
mapping $\lda$.\\
\begin{theorem}(\cite{Agarwal},\cite{Ortega})
Given the dynamical system \eqref{dynamical}, with $\lda: \cal{A} \mapsto \cal{A}$ and $\cal{A}$ $\subseteq \mathbb{R}^{m}$, we have the 
following:\\
If $\cal{A}$ is nonempty, convex and compact, and $\lda$ is a continuous mapping, then there exists
some $\mathbf{P}^{*}$ such that
\beq
\mathbf{P}^{*}=\lda(\mathbf{P}^{*}).
\enq
\end{theorem}

\begin{definition}
\label{preserving}
A function $\lda:\cal{A} \mapsto \cal{A}$ is locally gross direction preserving if for every $\mathbf{a} \in \cal{A}$ for which
$\lda(\mathbf{a}) \neq \mathbf{a}$, there exists $\delta> 0$ such that for every $\mathbf{b}, \mathbf{c} \in B(\mathbf{a},\delta)\cap A,$ the 
function satisfies 
\beq
\label{property}
 (\lda(\mathbf{b}) - \mathbf{b})^{T}(\lda(\mathbf{c})-\mathbf{c})\geq 0.
\enq
\end{definition}
 
\begin{theorem} (Ref.\cite{Hearing})
\label{preserve}
Given the dynamical system \eqref{dynamical}, with $\lda:\cal{A} \mapsto \cal{A}$ and $\cal{A}$ $\subseteq \mathbb{R}^{m}$, we have the 
following:\\
If $\cal{A}$ is a non-empty polytope in $\mathbb{R}^{m}$ and $\lda$ is locally gross direction preserving map then
there exists $\mathbf{P}^{*}$ such that
\beq
  \mathbf{P}^{*} = \lda(\mathbf{P}^{*}).
  \enq
\end{theorem}

\subsection{Updating strategies}

Fixed-point problems are typically solved by iterative methods, especially when one is interested in distributed algorithms \cite{Ortega}.
In fact, the mapping $\lda :\cal{A} \mapsto \cal{A}$ can be interpreted as an algorithm for finding such a fixed point. The degrees of freedom are in
the choice of the specific updating scheme among the components of vector $\mathbf{P}\in \cal{A}$, based on mapping $\lda$. Denoting by 
$\mathbf{P}=(\mathbf{P}_{1},\cdots,\mathbf{P}_{N})$ a partition of $\mathbf{P}\in \cA$, with $\mathbf{P}_{k}\in\mathbb{R}^{D}$ $\forall ~ k \in \left\{1,\cdots, N\right\}$, and assuming $\cal{A}$ $= {\cal{A}}_{1}\times\cdots\times {\cal{A}}_{N}$, with each ${\cal{A}}_{k}\subseteq\mathbb{R}^{D}$ and $\cA \subseteq \mathbb{R}^{m}$, where $m=ND$. We now give the definitions of some of the most common updating strategies for updating $\mathbf{P}_{1},\cdots,\mathbf{P}_{N}$ based on the mapping $\lda$.

\begin{definition}
\label{jacobi}
Jacobi Scheme :  All components $\mathbf{P}_{1},\cdots,\mathbf{P}_{N}$ are updated simultaneously, via the mapping $\lda$.
\end{definition}

\begin{definition}
\label{seidel}
Gauss-Seidel scheme :  All components $\mathbf{P}_{1},\cdots,\mathbf{P}_{N}$ are updated sequentially, one after the other, via the mapping $\lda$.
\end{definition}

\begin{definition}
\label{asynchronous}
Totally asynchronous scheme :  All components $\mathbf{P}_{1},\cdots,\mathbf{P}_{N}$ are updated in a totally asynchronous way, via the mapping $\lda$.
\end{definition}

For further details on updating strategies see \cite{Bertsekas,Ortega}.
\section{System Model}
\label{system model}
Consider a multicarrier data network with $N$ users, where each user has $D$ carriers over which 
it can transmit its data. We assume here that for every user there is a corresponding receiver, i.e., we have $N$ transmitter receiver pairs.
We further assume here that the carriers are sufficiently far apart so that the signal transmitted over a carrier does not 
interfere with the signals transmitted over other carriers. The received signal for the $k$-th user over the $l$-th carrier 
at its corresponding receiver after matched filtering can be represented as

\beq
\label{system}
 R_{kl} = \sqrt{p_{kl}h_{kl}}X_{kl} + {\displaystyle\sum_{i\neq k}}\sqrt{p_{il}g_{il}}X_{il} + w_{kl},
\enq

\noindent where $X_{kl}, p_{kl}, h_{kl}$ are the $k$-th  user's transmitted symbol, transmit power and  path gain respectively, for the $l$-th carrier, 
$X_{il}$ is the $i$-th user's transmitted symbol over its $l$-th carrier, $g_{il}$ is the co-channel path gain from user $i$ to the corresponding receiver of user $k$ and $w_{kl}$ is the complex Gaussian noise with
mean zero and variance $\sigma^{2}$. In the discussions below, we will assume that the channel undergoes slow fading with the path gain $h_{kl}$ 
and $g_{il}$ being exponentially distributed with parameter 1. We also assume that all the users choose their transmit symbol from the same constellation
and all the users have the same transmission rate. Under these assumptions and for a given set of
user's transmit vectors $\mathbf{P}_{1},\cdots, \mathbf{P}_{k}$, we define the utility for any user $k$ a function of $\mathbf{P}_{k}$ and $\mathbf{P}_{-k}$, where $\mathbf{P}_{k}$ is the power vector
in $\mathbb{R}^{D}$ and  $\mathbf{P}_{-k}$
is the set of all the user's transmit power vectors except the $k$-th user  in the same way as in \cite{Fahad}

\beq
\label{utility}
 U_{k}(\mathbf{P}_{k},\mathbf{P}_{-k}) = \frac{{\displaystyle\sum_{l = 1}^{D}} T_{kl}}{{\displaystyle{\sum_{l = 1}^{D}}p_{kl}}},
\enq

\noindent where $T_{kl}$ is the throughput achieved by the user $k$ over its $l$-th carrier, and is given by

\beq
 T_{kl} = R_{kl}f(\gma_{kl}),
\enq

\noindent where $\gma_{kl}$ is the received SINR and $f(\gma_{kl})$ represents the probability that a symbol transmitted  by the $k$-th user over its $l$-th carrier is received without error. We assume here that $f(\gma)$ is a continuous, increasing and S- shaped with the further property that 
$f(0)= 0$ and $f(\infty)=1$  Ref. \cite{analytic}. The utility function defined in \eqref{utility} has the unit of bits/joule.

\section{Game Theoretical Formulation}
\label{formulation}
We formulate the system design within the framework of game theory. Specifically, we consider a strategic noncooperative game, in which the players are
the transmitters and the pay off functions are the same as defined in \eqref{utility}. Each player $k$ competes against the others by choosing his transmit 
power vector $\mathbf{P}_{k}$ (i.e, his strategy) that maximizes his own utility $U_{k}(\mathbf{P}_{k},\mathbf{P}_{-k})$ in \eqref{utility}. We call this power
control game as $G_{N}$.
A solution of the game is called a Nash equilibrium when each user, given the strategy profiles of the others, does not get any increase in the 
utility by unilaterally changing his own strategy. Mathematically, the game can be expressed as $G_{N}$ is played by performing the following task
\beq
\label{game}
\max_{\bf{P}_k :\bf{P}_k \in {\cal{A}}_k} U_{k}(\bf{P}_k,\bf{P}_{-k})
\enq
$\forall ~ k \in \Omega$, where $\Omega= \{1,\cdots,N\}$ is the set of the players and ${\cal{A}}_{k}$ is the set of admissible strategies (the transmission power vectors) for player $k$, defined as
\begin{equation}
{\cal{A}}_{k} := \left\{\mathbf{P} \in [0,\text{P}_\text{max}]^{D}\right\},
\end{equation}
 \noindent where $\text{P}_\text{max}$ is the maximum transmit power allowed on each carrier. 
 
 \section{Nash Equilibrium For The Proposed Game}
 \label{best}
 \begin{definition}
A strategy profile $\mathbf{P}^{*}=(\mathbf{P}_{k}^{*})_{k\in \Omega}$ $\in$ ${\cal{A}}_{1}\times\cdots\times {\cal{A}}_{N}$ is
a Nash equilibrium point of the game $G_{N}$ if
\beq
 U_{k}(\mathbf{P}_{k}^{*},\mathbf{P}_{-k}^{*})\geq U_{k}(\mathbf{P}_{k},\mathbf{P}_{-k}^{*}), \forall ~ \mathbf{P}_{k} ~ \in ~ {\cal{A}}_{k}, \forall ~ k ~ \in \Omega.
\enq
\end{definition}

\begin{theorem} (Meshkati et al. Ref. \cite{Fahad})
\label{solution}
\noindent Given $k \in \Omega$ and $\mathbf{P_{-k}} \in {\cal{A}}_{-k}$, where ${\cal{A}}_{-k} = {\cal{A}}_{1}\times\cdots {\cal{A}}_{k-1}\times {\cal{A}}_{k+1}\times \cdots\times {\cal{A}}_{N}$, 
the solution to the problem defined in \eqref{game} is given by a set of power vectors $\mathbf{P}_{1}^{*},\cdots, \mathbf{P}_{k}^{*} $
which simultaneously satisfy

\beq
p_{kl} =
\begin{cases}
p_{kL_{k}}^{*} & \mbox{if } l = L_{k} \\
0         & \mbox{if } l \neq L_{k}
\end{cases}
\enq

\noindent where $L_{k} = \mbox{argmin}_{l}p_{kl}^{*}$ and $p_{kl}^{*}$ is the power required by the user $k$ over its $l$-th carrier to achieve $\gma^{*}$ which is the unique (positive) solution of $f(\gma)=\gma f^{'}(\gma)$ (for further details on the solution of $f(\gma)=\gma f^{'}(\gma)$ see Ref. \cite{analytic}).
\end{theorem}
It is shown in Ref.\cite{Goodman},  that if all the users transmit at the same rate then 
the optimal $\gamma^{*}$ achieved by all the users over all the carriers will be same. Theorem \ref{solution} suggests that at the Nash equilibrium point all the users will be transmitting only over the carrier which requires the least power to achieve $\gma^{*}$. From here on we will refer a carrier as the \emph{best carrier} for user $k$ if it requires the least amount of transmit power to achieve $\gma^*$ .\\

We now find the necessary conditions for carrier $l$ to be the best carrier for user $k$. We assume here that all the users get perfect feedback about the co-channel and noise interference from their respective receivers. Let $p_{kl}^{*}$ and $p_{ki}^{*}$ be the optimal power required by user $k$ on its $l$-th and $i$-th
carrier to achieve the optimal $\gamma^{*}$. Since the optimal $\gma^{*}$ achieved over all the carriers is same therefore,

\beq
\label{gain}
 \gma^{*} =  \frac{h_{kl}p_{kl}^{*}}{\displaystyle \sum_{j\neq k}g_{jl}p_{jl}^{*}+\sigma^{2}}=\frac{h_{ki}p_{ki}^{*}}{\displaystyle \sum_{j\neq k}g_{ji}p_{ji}^{*}+\sigma^{2}}.
\enq

\noindent Now if the carrier $l$ is the best carrier for the user $k$ then using \eqref{gain} the path gains must satisfy 
\beq
\label{cond.}
 \frac{h_{kl}}{h_{ki}}> \frac{\sigma^{2}+{\displaystyle\sum_{j\neq k}g_{jl}p_{jl}^{*}}}{\sigma^{2}+{\displaystyle\sum_{j\neq k}g_{ji}P_{ji}^{*}}}, ~ \forall ~ i \in [1:D],
\enq
where $h_{kl}$ and $g_{kl}$, $\forall$ $k$ $\in [1:N]$ and $\forall$ $l \in [1:D]$, are exponentially distributed
random variables with parameter $1$. \\

\section{Locally Gross Direction Preserving Property of the Best Carrier Strategy}
\label{direction preserving}
We prove the locally gross direction preserving property of the best carrier strategy for the two users, two carriers case. However, the proof can be easily generalized for any number of users and carriers. In the discussions below the map $\lda$ means the best carrier  strategy.\\

%
\noindent To prove the locally gross direction preserving property of $\lda$, we assume that every strategy vector 
$\mathbf{P}\in \cal{A}$ gets mapped to a unique strategy vector in $\cal{A}$ via the mapping $\lda$, i.e., at any time instant $n$ during the update process,
\beq
\label{assumption}
 p_{kl}^{*}(n)\neq p_{km}^{*}(n),\hspace{2mm} \forall ~ k\in \Omega ~ \text{and} ~ \forall~ l\neq m. 
\enq
For further discussion on \eqref{assumption} see Section \ref{simulation}. Let $\bP = [p_{11}~p_{12}~p_{21}~p_{22}] \in {\cal{A}}$ be a vector in $\mathbb{R}^{4}$, and suppose that
\beq
\label{map}
 \lda(\mathbf{a}) = \left [p^{*}_{11}~0~p^{*}_{21}~0\right].
 \enq
 For \eqref{map} to be true the channel gains of user 1 and user 2 must satisfy \eqref{cond.}, i.e.,
 
\begin{align}
 \frac{h_{11}}{h_{12}} & > \frac{\sigma^{2}+g_{21}a_{21}}{\sigma^{2}+g_{22}a_{22}},\\
  \frac{h_{21}}{h_{22}} & > \frac{\sigma^{2}+g_{11}a_{11}}{\sigma^{2}+g_{12}a_{12}}.
 \end{align}

\noindent Therefore, for some $\epsilon_{1}>0$, $\epsilon_{2}>0$,

\begin{align}
 \frac{h_{11}}{h_{12}} &>\frac{\sigma^{2}+g_{21}a_{21}}{\sigma^{2}+g_{22}a_{22}} + \epsilon_{1}, \\
 \frac{h_{21}}{h_{22}} &> \frac{\sigma^{2}+g_{11}a_{11}}{\sigma^{2}+g_{12}a_{12}} + \epsilon_{2}.
 \end{align}
 
 \noindent Let $\epsilon=\min\{\epsilon_{1}, \epsilon_{2}\}$. Then $\exists$ $\delta>0$, (depending upon $\epsilon$) s.t. $\forall$ $\mathbf{b}$
$\in$ $\cal{A}\cap\mathbf{B}(\mathbf{a},\delta)$ the following holds

\begin{align}
 \frac{h_{11}}{h_{12}} &>\frac{\sigma^{2}+g_{21}b_{21}}{\sigma^{2}+g_{22}b_{22}} + \epsilon,\\
\frac{h_{21}}{h_{22}}  &> \frac{\sigma^{2}+g_{11}b_{11}}{\sigma^{2}+g_{12}b_{12}} + \epsilon.
\end{align}

The above argument shows that the nearby strategy vectors get mapped to the same carriers by the best carrier response strategy. Now let 
$\lambda(\mathbf{b}) = [p^{\prime}_{11}\hspace{1mm}0\hspace{1mm}p^{\prime}_{21}\hspace{1mm}0]$, where

\begin{align}
\label{existence1}
 \frac{h_{11}p^{\prime}_{11}}{\sigma^{2}+g_{21}b_{21}} & = \gamma^{*}, \\
 \label{existence2}
\frac{h_{21}p^{\prime}_{21}}{\sigma^{2}+g_{11}b_{11}} & =\gamma^{*}.
\end{align}

\noindent Now since $\mathbf{b}$ $\in \cal{A}\cap\mathbf{B}(\mathbf{a},\delta)$, using \eqref{existence1} and \eqref{existence2} it follows that $\lda(\mathbf{b})$ $\in \mathbf{B}(\lda(\mathbf{a}),\delta_{1})$ for small $\delta_{1}$ (depending on $\delta$).
Since $\mathbf{a}$ is arbitrary therefore for every $\mathbf{a}\in A$ there exists a $\delta$ neighborhood such that $\forall$ $\mathbf{b}, \mathbf{c} \in B(\mathbf{a},\delta)\cap\cal{A}$, \eqref{property}
is satisfied.  Hence, the best carrier strategy is a locally gross direction preserving map. Thus, using Theorem \ref{preserve} it follows that the noncoperative power control game has a Nash equilibrium point with probability $1$.

\section{A greedy strategy to achieve the Nash equilibrium point}
\label{algo}
In this section we prove the convergence of the best carrier strategy for two users and two carriers case.  In the discussions below, we assume that at time $n = 1$, both the users start with any arbitrary transmit power vector and then they update there transmit power vector using Gauss-Seidel  scheme, definition \ref{seidel}. We define the following notations which will be needed in the subsequent proofs.
 \begin{align*}
 (1,2) &\to \mbox{user 1 is on carrier 1 and user 2 is on carrier 2},\\
(2,1) &\to \mbox{user 1 is on carrier 2 and user 2 is on carrier 1},\\  
(12, ) & \to \mbox{both the users are on carrier 1},\\
(  ,12) & \to \mbox{both the users are on carrier 2}, \\ 
p_{kl}^{*}(n) & \to \mbox{optimal transmit power of any user $k$ over its $l$-th carrier at time n},\\ 
p_{kl}^{'} & \to \mbox{optimal transmit power of any user $k$ over its $l$-th carrier at the Nash equilibrium point}.
\end{align*}

\subsection{The Case in which $(12, )$ is the Nash Equilibrium Point}
In this case the received SINR $\gma^{*}$ and the transmitted powers of user 1 and user 2 satisfy the following set of inequalities at the Nash equilibrium point.
\begin{align}
\label{pow1}
 p_{11}^{'} & < \frac{\sigma^{2}}{g_{11}}\left(\frac{h_{21}}{h_{22}}-1\right),\\
 \label{pow2}
 p_{21}^{'} & < \frac{\sigma^{2}}{g_{21}}\left(\frac{h_{11}}{h_{12}}-1\right),\\
 \gma^{*} & < \min \left\{\frac{h_{11}}{g_{11}}\left(\frac{h_{21}}{h_{22}}-1\right), \frac{h_{22}}{g_{21}}\left(\frac{h_{21}}{h_{22}}-1\right)\right\}.
\end{align}

\noindent Let $\eta_{1}:= \frac{\sigma^{2}}{g_{11}}\left(\frac{h_{21}}{h_{22}}-1\right)$ and $\eta_{2} := \frac{\sigma^{2}}{g_{21}}\left(\frac{h_{11}}{h_{12}}-1\right).$ 
 At any time $t$, if $p_{11}^{*}(n) < \eta_{1}$ and $p_{21}^{*}(n) < \eta_{2}$ then 
carrier one will be the best carrier for both the users. In the discussions
below, to prove the convergence of the best carrier strategy we make use of the fact that
\beq
\label{iff}
 p_{11}^{*}(n)<p_{11}^{'}\Leftrightarrow p_{21}^{*}(n)<p_{21}^{'},
\enq
\noindent The if and only if condition of \eqref{iff} follows from the Nash equilibrium conditions mentioned in \eqref{pow1} and \eqref{pow2}. 

\subsubsection{Starting from (12, )}
 
 If user $1$ (user $2$) updates first and jumps to the second carrier then in the next iteration user $2$ (user$1$) remain on the first carrier with the optimal power $P_{21}^{*}(3)$ ($P_{11}^{*}(3)$) $<P_{21}^{'} $ ($P_{11}^{'}$), as a result user $1$ (user $2$) returns to carrier one in the next
iteration. From here on both the users stay on the same carrier until the Nash equilibrium is achieved.\\

If user $1$ updates first and remains on the first carrier with its updated power $p_{11}^{*}(1)$ satisfying   
\begin{equation}
p_{11}^{'}<p_{11}^{*}(2)<\eta_{1},
\end{equation}

\noindent then in the next iteration user $2$ will remain on the first carrier with its updated power $p_{21}^{*}$. Now suppose that after the $n$-th iteration 
the updated power of user $1$ satisfies $p_{11}^{*}(n)>\eta_{1}$, then in the $(n+1)$-th iteration user $2$ will switch to the second carrier.
Therefore, in the $(n+2)$-th iteration user $1$ will remain on the first carrier with the updated power $p_{11}^{*}(n+2)<p_{11}^{'}$. Hence,
in the $(n+3)$-th iteration user $2$ will switch back to the carrier one. Thus, after $(n+3)$-th iteration both the users remain on carrier one until Nash equilibrium is achieved. Similar argument holds for the case when user $2$ starts first.\\

If user $1$ updates first and remains on the first carrier with its updated power $p_{11}^{*}(2)$ satisfying 
\begin{equation}
 p_{11}^{*}(2)<p_{11}^{'},
\end{equation}
\noindent then in the next iteration user $2$ also remains on the same carrier with its updated power $p_{21}^{*}(3)$ satisfying
\begin{equation}
  p_{21}^{*}(3)<p_{21}^{'}.
\end{equation}
Thus, after the second iteration onwards both the users remain on the same carrier until Nash equilibrium is achieved. Similar argument holds for the case when user $2$ starts first.

\subsubsection{Starting From $(1,2)$}

From \eqref{pow1} and \eqref{pow2}, it follows that carrier one is the best carrier for both the user. 
If the user $1$ updates first then it will remain on first carrier with $p_{11}^{*}(2)<\eta_{1}$. 
Hence user $2$ in the third iteration will switch to the first carrier. From here on both the users stay on the first carrier until Nash equilibrium is achieved.\\

If user $2$ updates first and stays on the second carrier, then we are back to the case we just described. If user $2$ jumps to the first carrier 
then we are back to the case of (12, ). On the other hand, if user $2$ jumps to the first carrier and user $1$ jumps to the second carrier in the second iteration, 
then user $2$ will remain there in third iteration too with $p_{21}^{*}(4)< p_{2}^{'}$. Hence, in the fourth iteration user $1$ will jump back to the 
first carrier. Thus, from the fifth iteration onwards both the users stay on the first carrier until Nash equilibrium is achieved.

\subsubsection{Starting From $( ,12)$}
The argument is similar to $(12, )$ due to symmetry.

\subsubsection{Starting From $(2,1)$}
The argument is similar to the case $(1,2)$ due to symmetry.

\subsubsection{Starting From $( ,12)$}
The argument is similar to $(12, )$ due to symmetry.
\subsubsection{Starting From $(2,1)$}

The argument is similar to the case $(1,2)$ due to symmetry.
\subsection{The Case in which $(1,2)$ is the Nash Equilibrium}
In this case the channel gains and the transmitted powers of user $1$ and user $2$ satisfy the following set of inequalities at the Nash equilibrium point.
\begin{align}
 \frac{h_{11}}{h_{12}} &>\frac{\sigma^{2}}{\sigma^{2}+g_{22}p_{22}^{'}},\\
 \frac{h_{22}}{h_{21}} &>\frac{\sigma^{2}}{\sigma^{2}+g_{11}p_{11}^{'}},\\
 p_{11}^{'} &= \frac{\gamma^{*}\sigma^{2}}{h_{11}},\\
 p_{22}^{'}& = \frac{\gamma^{*}\sigma^{2}}{h_{22}}.
\end{align}

\subsubsection{Starting from $(12, )$}

If user 1 updates first and stays on the first carrier then the updated power $p_{11}^{*}(2)$ of the user one will be greater than $p_{11}^{'}$.
Thus, user 2 will jump to carrier two in the next iteration with $p_{22}^{*}(3)=p_{22}^{'}$. 
As a result, user $1$ stays on the first carrier in the next iteration with $p_{11}^{*}(4)=p_{11}^{'}$. Hence, Nash equilibrium is achieved.\\

If user $2$ updates first, it will jump to the second carrier with $P_{22}^{*}(2) = p_{22}^{'}$. As a result, user $1$ will stay on the first carrier with 
$p_{11}^{*}(3)= p_{11}^{'}$. Thus, Nash equilibrium is achieved. \\

If user $1$ updates first but jumps on the second carrier, then because we are assuming that (1,2) is the Nash equilibrium, user $2$ will also jump to 
the second carrier in the next iteration. As a result, user $1$ will jump on carrier one with $p_{11}^{*}(4) = p_{11}^{'}$ and user $2$ will stay on
carrier two with $p_{22}^{*}(5) = p_{22}^{'}$. Thus, Nash equilibrium is achieved.

\subsubsection{Starting from $(1,2)$}
If user $1$ (user $2$) updates first, it will stay on the first (second) carrier with $p_{11}^{*}(2) =  p_{11}^{'}$ ($p_{22}^{*}(2) =  p_{22}^{'}$).
As a result user $2$ (user $1$) stays on the second (first) carrier. Thus, Nash equilibrium is achieved.

\subsubsection{Starting from $(2,1)$}
If user $1$ (user $2$) updates first and jumps to the first (second) carrier, then user user $2$ (user $1$) will jump to the second (first carrier) in the 
second iteration with $p_{22}^{*}(3) = p_{22}^{'}$ ($p_{11}^{*}(3) = p_{11}^{'}$). Hence; in the third iteration
the power of user $1$ (user$2$) is $p_{11}^{*}(4) = p_{11}^{'}$ ($p_{22}^{*}(4) = p_{22}^{'}$). Thus, Nash equilibrium is 
achieved. \\

If user $1$ (user $2$) updates first but it stays on the second (first) carrier, then because we are assuming that (1,2) is the Nash equilibrium,
user $2$ (user $1$) will jump to carrier two (one). As a result, user $1$ (user $2$) jumps to the first (second carrier). Thus, user $2$ (user $1$)
will stay on the second (first) carrier and equilibrium is achieved.\\

The other cases of Nash equilibrium follow straightforwardly using the same argument as above because of the symmetry of the problem.

\section{Simulation results}
\label{simulation}

We now present the simulation results of the power control game for the two users and two carriers case. In the simulation we assume that that channels gains are exponentially distributed and the noise is Gaussian distributed with variance $1$. We also assume that the maximum transmission power allowed over all the carriers is 1000 units and at time instant $n = 1$, both the users transmit on a randomly
chosen carrier using 100 units of transmit power to start the iterative process. We did not find that this choice affected the conclusions
in any way. We check the convergence of the best carrier strategy for $10$ million power control games. In each of these simulation results, we did not find a case where the assumption in \eqref{assumption} is not valid. Figure \ref{fg} below shows a typical convergence of the transmit power vector of user 1 and user 2 to the Nash equilibrium point. \\

\begin{figure}[h!]
  \centering
    \includegraphics[width=0.8\textwidth]{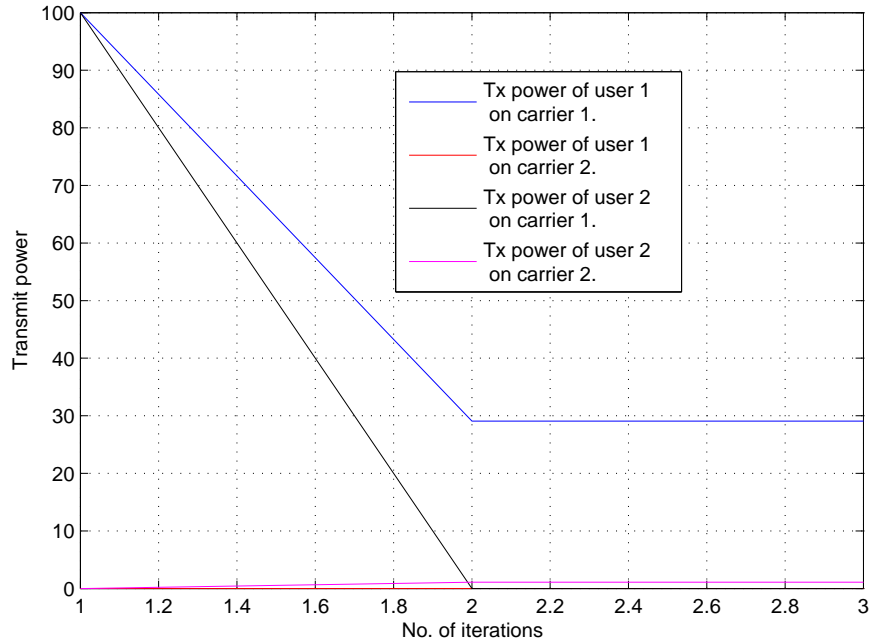}
     \caption{Convergence of the best carrier strategy for two users and two carriers case.}
     \label{fg}
\end{figure}

\section{Conclusions And Acknowledgment}
\label{con}
We presented a game theoretic model to study the management of transmission power in a multiuser multicarrier data network in the presence of co-channel interference. We derived necessary conditions on the path gains when the Nash equilibrium exists. We also showed that the best carrier strategy is a locally gross direction preserving map. We give a greedy strategy to achieve the Nash equilibrium point. Finally, simulation results also show that the best carrier strategy is a locally gross direction preserving map.

This work was done under the supervision of Dr. Rahul Vaze. Dr. Vaze has
observed that arguments similar to those used in this paper to show the existence
of Nash equilibrium in the multicarrier setting can be used to show the existence
of Nash equilibrium in the MIMO setting Ref. \cite{vaze}.

\bibliographystyle{ieeetr}
\bibliography{master}

\end{document}